\documentclass[12pt]{iopart}
\usepackage{graphicx}
\begin{document}

\title[The angle of repose of spherical grains...]{The angle of repose of spherical grains in granular Hele-Shaw cells: A
molecular dynamics study }

\author{Hamed Maleki}
\address{Physics department, University of Birjand,
Birjand 97175-615, Iran}

\author{Fatemeh Ebrahimi \footnote[1]{corresponding author}}
\address{Physics department, University of Birjand,
Birjand 97175-615, Iran} \ead{febrahimi@birjand.ac.ir}

\author{Ehsan Nedaaee Oskoee}
\address{Physics department, Institute for Advanced Studies in Basic
Sciences (IASBS), Zanjan 45195-1159, Iran}

\begin{abstract}
We report the results of three dimensional molecular dynamic
simulations on the angle of repose of a sandpile formed by
pouring mono-sized cohesionless spherical grains into a granular
Hele-Shaw cell. In particular, we are interested to investigate
the effects of those variables which may impact significantly on
pattern formation of granular mixtures in Hele-Shaw cells. The
results indicate that the frictional forces influence remarkably
the formation of pile on the grain level. Furthermore, We see
that increasing grain insertion rate decreases slightly the angle
of repose. We also find that in accordance with experimental
results, the cell thickness is another significant factor and the
angle of repose decays exponentially by increasing the cell
thickness. It is shown that this effect can be interpreted as a
cross-over from two to three dimensions. In fact, using grains
with different sizes shows that the behavior of the angle of
repose when both size and cell thickness are varied is controlled
by a scaled function of the ratio of these two variables.

\end{abstract}

\maketitle
\section{introduction}
The formation of a sandpile is an everyday life experience which
may occur in several ways, including pouring~\cite{gh},
discharging~\cite{dis1,dis2}, failing~\cite{lee1,spf}, tilting
and rotating~\cite{spr1,spr2}. This gravity-driven physical event
is relevant to many practical applications and laboratory
processes such as avalanches~\cite{ava1,ava2,ava3} and pattern
formation in granular mixtures~\cite{mhks,mcs}. It is also
related to the paradigm of self organized criticality~\cite{soc1}
which has itself a wide realm of applicability to a diverse range
of physical phenomena~\cite{soc2}.

The angle of repose of a sandpile is a simple concept that
characterizes the behaviour of granular materials on macroscopic
scale. A sandpile has an inclined free surface that does not
flow. When grains are added to the sandpile the slope of this
free surface increases until, it exceeds a threshold value. At
this point, the pile does not support new grains and releases
many grains in an avalanche reducing  the slope to the angle of
repose, $\theta_r$. One of the most important questions in the
study of piling of grains is how to determine the relationship
between the angle of repose and the relevant variables, i.e.
material properties, boundary conditions, and the model
interactions at the grain level. Experiments have shown that
$\theta_r$ depends on the shape, density, and size distribution
of the grains, as well as humidity, formation history and
geometrical boundary conditions~\cite{gh,dis1,dis2}.

Granular materials are ensembles of discrete particles, and as
for any macroscopic system, the total number of modes are huge.
What we see on macroscopic scale, hence, is a manifestation of
the collective behaviour of a large assembly of macroscopic grains
that interact with each other (and maybe with the container
walls), through collisions and friction. In this regard, the
molecular dynamics (MD) simulation technique provides perhaps the
most realistic approach for modeling the properties of granular
materials. The present work is focused on such a numerical study
of the angle of repose of a heap formed by pouring dry,
mono-sized spherical grains into a granular Hele-shaw cell. Some
of the experimental aspects of the phenomenon have been reported
by Grasselli and Herrmann~\cite{gh}. However, as far as we know,
there has been no three-dimensional MD study of the angle of
repose in Hele-Shaw cell, except the one by Zhuo et
al~\cite{xyz}, which studied the formation of sandpiles resulting
from discharging materials by using a different model
interaction. Regarding to the classification made by Grasselli
and Herrmann~\cite{gh}, what they have estimated is the outflow
angle of the heap which is less than the heap angle, the subject
of current research.

In this paper, we report a detailed numerical study of the
formation of a sandpile and the dependence of its angle of repose
on the relevant variables. We are specifically interested to
determine how $\theta_r$ changes with variables like the size of
particles, the rate of mass injection, the density of grains, the
thickness of the container, and wall-particle and
particle-particle static friction coefficients. These are the key
parameters which may impact significantly on pattern formation of
granular mixtures in Hele-Shaw cells~\cite{mbsw,kop}. To answer
such questions, we performed extensive MD simulations in three
dimension on model systems by using a new MD scheme developed by
Silbert et al~\cite{silp}, which is described briefly in the next
section. The results of our simulations are presented in section
$3$, followed by the main conclusions at section $4$.
\section{SIMULATION METHODOLOGY}
\begin{figure}
\begin{center}
\includegraphics[width=3 in]{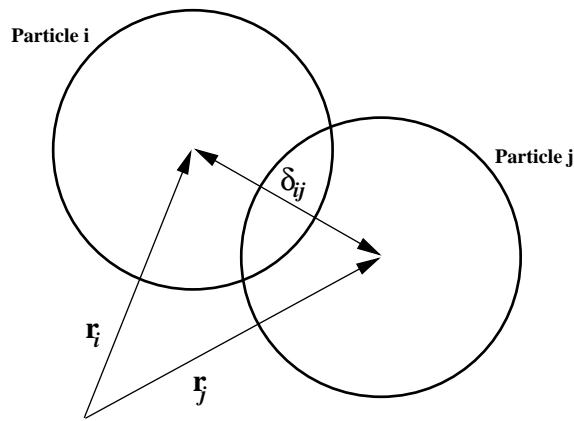}

\caption{Two-dimensional illustration of two grains $\textit{i}$
and $\textit{j}$ in contact and position vectors $\textbf{r}_{i},
\textbf{r}_{j}$, respectively, with compression  and overlap
$\delta_{i,j}$.} \label{fig1}
\end{center}
\end {figure}

As mentioned earlier,  our simulations are based on a Distinct
Element Method (DEM) scheme, originally proposed by Silbert et
al. The method has been discussed elsewhere in
details~\cite{silp}. For the sake of clarity, we review the most
important aspects of it here. According to this model, the
spheres interact on contact through a linear spring dashpot or
Hertzian interactions in the normal and tangential directions to
their lines of centers. In a gravitational field $g$, the
translational and rotational motions of grain $i$ in a system at
time $t$, caused by its interactions with neighboring grains or
walls, can be described by Newton second law, in term of the
total force and torque as the following equations:
\begin{equation}
\textbf{F}^{tot}_{\textit{i}}=m_{\textit{i}}\textbf{g}+\sum\left(
\textbf{F}_{n_{\textit{i},\textit{j}}}+\textbf{F}_{t_{\textit{i},\textit{j}}}\right)
\end{equation}
\begin{equation}
\tau^{tot}_{\textit{i}}= -\frac{1}{2}\sum\textbf{r}_{\textit{i},
\textit{j}}\times\textbf{F}_{t_{\textit{i}, \textit{j}}}
\end{equation}
Where $m_{i}$,$ \textbf{r}_{i,j}=\textbf{r}_{i} - \textbf{r}_{j}$
are respectively, the mass of grain \textit{i} and the relative
distance between grains \textit{i} and \textit{j}. For two
contacting grains \textit{i}, \textit{j} at positions
$\textbf{r}_{i}$, $\textbf{r}_{j}$ with velocities
$\textbf{v}_{i}, \textbf{v}_{j}$ and angular velocities
$\omega_{i}, \omega_{j}$ the force on grain \textit{i} is
computed as follows: the normal compression $\delta_{i,j}$ is
(Fig.\ref{fig1})
\begin{equation}
\delta_{i,j}={d}-{r}_{\textit{i,j}}
\end{equation}

and relative normal velocity $\textbf{v}_{n_{i,j}}$ and relative
tangential velocity $\textbf{v}_{t_{i,j}}$ are given by
\begin{equation}
\textbf{v}_{n_{i,j}}=(\textbf{v}_{i,j}.\textbf{n}_{i,j})
\textbf{n}_{i,j}
\end{equation}
\begin{equation}
\textbf{v}_{t_{i,j}}=\textbf{v}_{i,j}-\textbf{v}_{n_{i,j}}-\frac{1}{2}(\omega_{i
}+\omega_{j}) \textbf{r}_{i,j}
\end{equation}
Where $\textbf{n}_{i,j}=\frac{\textbf{r}_{i,j}}{r_{i,j}}$ with
$r_{i,j}=\vert \textbf{r}_{i,j}\vert$ and
$\textbf{v}_{i,j}=\textbf{v}_{i}-\textbf{v}_{j}$. The rate of
change of elastic tangential displacement $\textbf{u}_{t_{i,j}}$,
set to zero at the initial of a contact, is given by \cite{silp}
\begin{equation}
\frac{d\textbf{u}_{t_{i,j}}}{dt}=\textbf{v}_{t_{i,j}}-\frac{(\textbf{u}_{t_{i,j}
}.\textbf{v}_{i,j}) \textbf{r}_{i,j}}{r_{i,j}^{2}}
\end{equation}

The second term in Eq.(6) arises from the rigid body rotation
around the contact point and insures that $\textbf{u}_{t_{i,j}}$
always lies in the local tangent plan of contact. In Eqs. (1) and
(2), normal and tangential forces acting on grain \textit{i} are
given by
\begin{equation}
\textbf{F}_{n_{i,j}}=f(\frac{\delta_{i,j}}{d})(k_n\delta_{i,j}\textbf{n}_{i,j}-
\gamma_{n}m_{eff}\textbf{v}_{n_{i,j}})
\end{equation}
and
\begin{equation}
\textbf{F}_{t_{i,j}}=f(\frac{\delta_{i,j}}{d})(k_{t}\textbf{u}_{t_{i,j}}-\gamma_
{t}m_{eff}\textbf{v}_{t_{i,j}})
\end{equation}
Where $k_{n,t}$ and $\gamma_{n,t}$ are elastic and viscoelastic
constants respectively and $m_{eff}=m_{i}m_{j}/(m_{i}+m_{j})$ is
the effective mass of spheres with masses $m_{i}$ and $m_{j}$.
The corresponding contact force on grain \textit{j} is simply
given by Newton’s third law, i.e.,
$\textbf{F}_{i,j}=-\textbf{F}_{j,i}$ . For spheres of equal mass
$\textit{m}$, as is the case in our system, $m_{eff}=m/2$;
$f(x=1)$ for the linear spring dashpot (Hookian) model, and
$f(x)=\sqrt{x}$ for Hertzian contacts with viscoelastic damping
between spheres \cite{silp}.

Static friction is implemented by keeping track of the elastic
shear displacement throughout the lifetime of a contact. The
static yield criterion, characterized by a local particle
friction coefficient $\mu$, is modeled by truncating the
magnitude of $\textbf{u}_{t_{i,j}}$ as necessary to satisfy
$\vert\textbf{F}_{t_{i,j}}\vert<\vert\mu\textbf{F}_{n_{i,j}}\vert$.
Thus the contact surfaces are treated as “sticking” when
$\vert\textbf{F}_{t_{i,j}}\vert<\vert\mu\textbf{F}_{n_{i,j}}\vert$,
and as “slipping” when the yield criterion is satisfied
\cite{Landry, Sun}.

\section{RESULTS AND DISCUSSION}
\begin{figure}
\begin{center}
\includegraphics[width=2.5 in , height=3 in]{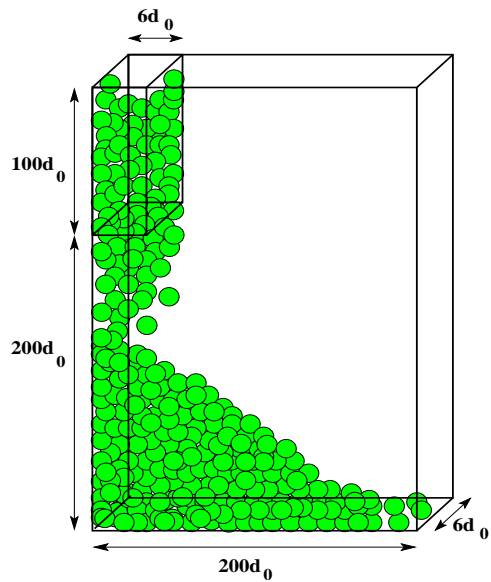}

\caption{Geometry of cell used in simulations.} \label{fig2}
\end{center}
\end {figure}

In this section we present the results of our extensive molecular
dynamics (MD) simulations in three dimensions on model systems of
$N$ mono-disperse, cohesionless and inelastic spheres of diameter
$d$ and mass $m$. The system is constrained by a rectangular box
with fixed rough walls and boundary conditions and free top
surface, as in Fig.\ref{fig2}. A simulation was started with the
random generation of spheres without overlaps from top and left
corner of container, followed by a gravitational settling process
to form a stable heap. The results are given in non-dimensional
quantities by defining the following normalization parameters:
distance, time, velocity, forces, elastic constants, and stress
are, respectively measured in units of \textit{d},
$t_{0}=\sqrt{d/g}$, $v_{0}=\sqrt{dg}$, $F_{0}=mg$, and
$k_{0}=mg/d$. All data was taken after the system had reached the
steady state. Because of the complexity of the model, there are a
wide range of parameters that affect the results of computation.
However, we usually investigate the effect of a single variable
varying in a certain range while other variables are fixed to
their base values as listed in table \ref{table1}.

All the cases were simulated in three dimensions using a molecular
dynamics code for granular materials LAMMPS \cite{silp, Lammps}.
The equations of motion for the translational and rotational
degrees of freedom are integrated with either a thirdorder Gear
predictor-corrector or velocity-Verlet scheme \cite{Plimpton}.
The angle of repose could then be determined from the surface
profile of the heap.

\subsection{The evolution of the pile}
We start from an empty vertical cell composed of two parallel
plates separated by an spacer with a variable width $w$. A
granular heap is formed in the cell by pouring mono-sized
spherical grains on the bottom plate. These grains are released
from a small box located on the top-left of the cell, as shown in
Fig.\ref{fig1}. The rate of material insertion $R$ can be varied
by changing the box size. A pile with well defined shape starts
to form as the number of the added grains growth. The angle of
repose could then be determined from the surface profile of the
heap.

\begin{table}
\caption{\label{table1}Basic computational parameters}
\begin{indented}
\item[]\begin{tabular}{@{}ll}
\br
Parameters&{\rm system conditions}\\
\mr
\verb"Number of Particles "$(N)$&$40,000$\\
\verb"Particle Size "$(d)$&$1d_0$ \\
\verb"Particle density "$(\rho)$&$2.4 (m_0/d_0^3)$ \\
\verb"Particle friction coefficient " $(\mu_{p})$&$0.5$\\
\verb"Wall friction coefficient "$(\mu_{w}) $&$0.5$\\
\verb"Particle normal stiffness coefficient "$(k_n )$&$2\times10^3(k_0)$\\
\verb"Particle tangential stiffness coefficient "$(k_t )$&$\frac{2}{7}k_n$\\
\verb"Particle normal damping coefficient "$(\gamma_n )$&$50/(t_0)$\\
\verb"Particle tangential damping coefficient "$(\gamma_t )$&$0(t_0^{-1})$\\
\verb"Wall normal stiffness coefficient "$(k_n )$&$2\times10^3(k_0)$\\
\verb"wall tangential stiffness coefficient "$(k_t )$&$\frac{2}{7}k_n$ \\
\verb"wall normal damping coefficient "$(\gamma_n )$&$50/(t_0) $ \\
\verb"Particle normal damping coefficient "$(\gamma_t)$&$50/(t_0) $ \\
\verb"Time step increment "&$2\times10^{-3}$\\

\br
\end{tabular}
\end{indented}
\end{table}

\begin{figure}
\begin{center}
\includegraphics[width=4 in]{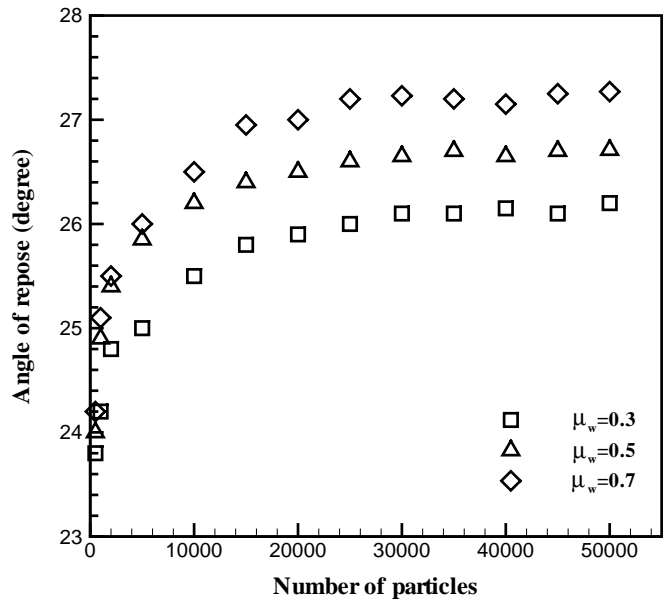}

\caption{Angle of repose as a function of number of grains from
Hookin model.} \label{fig3}
\end{center}
\end{figure}

The formation of a granular heap at the macroscopic level is a
complicated phenomenon arises as the result of surface avalanches
and also a kink mechanism~\cite{alonso2}. At the grain level, it
is a consequence of energy dissipation. In addition to frictional
forces, the particle-particle inelastic collisions consumes the
translational and rotational energy of a falling grain.
Eventually, the grain will stick to a position where its energy
is less than the minimum value required to overcome the potential
barrier created by frictional forces~\cite{alonso1}. At the
earlier stage of the pile formation, the number of grains are
small. Therefore, the falling grains can move further and the
angle of repose is less than the one of a larger pile. By adding
more grains to the pile the angle of repose growth to its
asymptotic value, where the energy gain in the gravitational
field is balanced by the dissipating effects. In Fig.\ref{fig3}
the evolution of the sandpile or equivalently, the evolution of
$\theta _r$ has been plotted for two values of $\mu_w$ against
$N$, the total number of released grains. As seen from the
figure, the angle of repose grows linearly for smaller values of
$N$ and then  evolves quickly into a steady state, characterized
by an asymptotic value of $\theta_r$ at larger values of $N$. For
the set of parameters used in Fig.\ref{fig3}, the angle of repose
is almost constant for $N>25,000$. For the rest of the work we
have used $N=40,000$ grains in all simulations to make sure that
the angle of repose has reached its asymptotic values as shown in
Fig.\ref{fig3}.

We also note that at earlier stage of heap formation the
difference between $\theta(N)$'s is not significant for different
values of $\mu_w$, as seen in Fig.\ref{fig3}. This maybe explains
the existence of the so called dead-zone~\cite{mcs} in
stratification of bi-dispersed mixtures. It is known that the
difference between the angle of repose of two species is the key
factor which together with the difference between the grain sizes
determines the type of patterns formed in the cell~\cite{mhks}.
If the difference between $\theta _r$ is negligible, e.g., when
the number of grains is small, the only factor that affects
pattern formation is the size. As such, for small values of $N$,
the grain segregate according to their size and a dead-zone forms.

\subsection{Effect of mass flow rate}
\begin{figure}
\begin{center}
\includegraphics[width= 4 in]{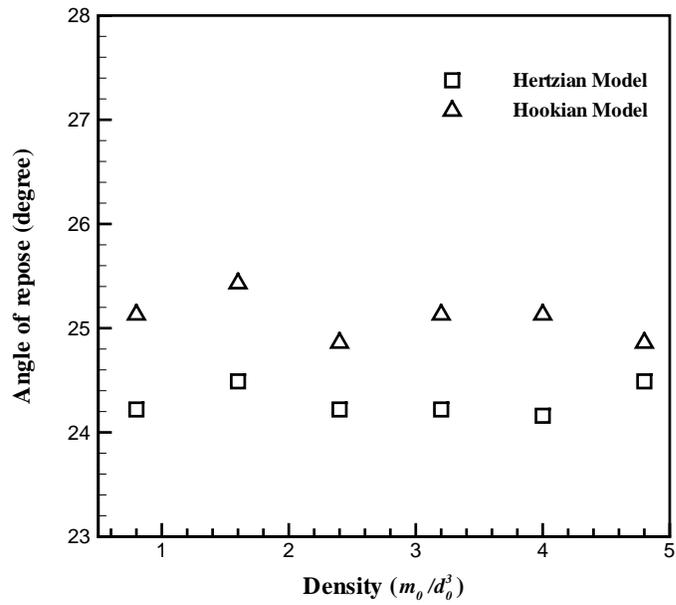}
\end{center}
\caption{The angle of repose against density from Hookian and
Hertzian Models} \label{fig4}
\end {figure}

A simple way of increasing the rate of mass flow into the cell is
increasing the density of each grain, while its diameter is
fixed. From Fig.\ref{fig4} one can see that increasing the mass of
individual grains causes a slight fluctuation, at most $0.3^o$
around the mean values in both linear and Hertzian model. In
fact, there is no hint of a dependence of $\theta_r$ on the value
of grain's mass in our simulations in the range of parameters we
used. This conclusion is in agreement with some experimental
observations~\cite{mhks}.

\begin{figure}
\begin{center}
\includegraphics[width=4 in]{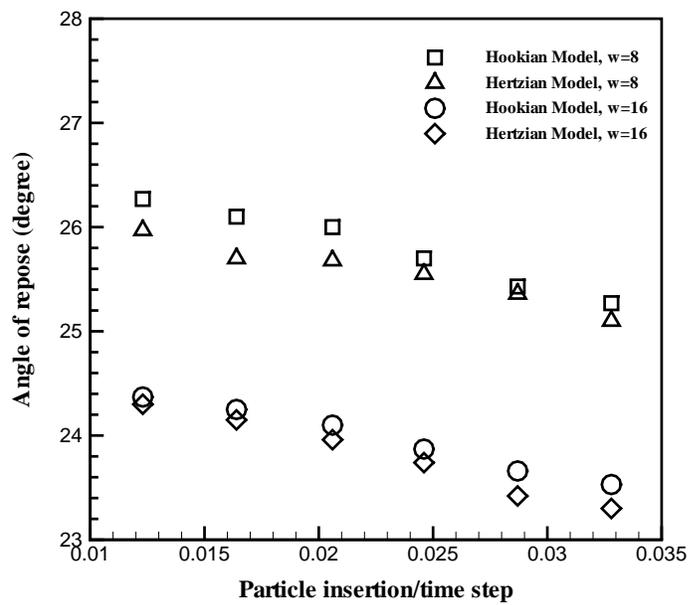}
\end{center}
\caption{Angle of repose as a function of grain insertion from
Hookian and Hertzian models for two different container
thickness.} \label{fig5}
\end{figure}

A more interesting way of increasing mass flow rate is to increase
$R$, the number of grains falling down on the heap at each time
step. It has been found that the wavelength of the layers in
granular stratification is an increasing function of this quantity
~\cite{mbsw}. Moreover, above a critical flow rate $R_c$, the
stratification pattern abruptly disappear~\cite{kop}. Our
simulations shown in Fig.\ref{fig5} indicate that, increasing the
rate of grain insertion $R$, decreases the angle of repose
slightly. The change in $\theta_r$ is small, about a degree after
increasing $R$ three times. Such a slight change in the value of
angle of repose when the rate of grain insertion was varied has
been reported by Grasselli and Herrmann~\cite{gh}.

Comparing the two ways of increasing $R$, we conclude that in the
formation of a sandplie, what does really matter is not the mass
itself, but the number of grains that hit the pile in a given
time. This effect may be attributed to the occurrence of larger
avalanches when the rate of grains falling upon top of the heap
becomes larger.

\subsection{Effect of cell thickness}
The effect of variation in the size of the gap between walls
container on the piling of grains has been the subject of both
experimental and numerical studies with the general conclusion
that increasing the cell thickness induces an exponential decay
in the angle of repose~\cite{gh}
\begin{equation}
\theta_r(w)=\theta_{\infty}(1-\alpha\: e^{-w/l})
\end{equation}
where $\alpha$ is a constant depending on the grain properties
and $l$ is a characteristic length representing the scale over
which the walls affect the piling of grains. The parameter $w$ is
also a determining factor which affects drastically pattern
formation of bi-dispersed granular mixtures in Hele-Shaw cells. A
nonlinear decrease in the wavelength of the stratification
pattern with cell thickness $w$ has been reported ~\cite{mbsw}.
And similar to $R_c$, there is a critical cell thickness $w_c$
above which the layered pattern abruptly vanishes~\cite{kop}.

\begin{figure}
\begin{center}
\includegraphics[width=4 in]{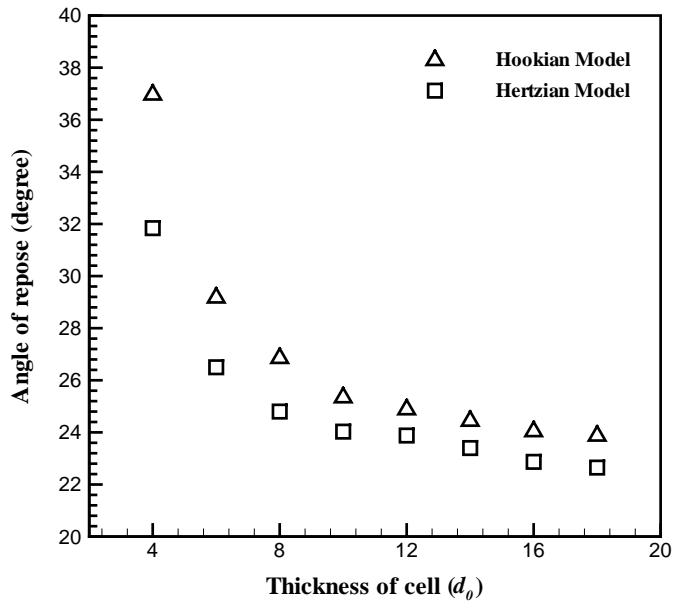}
\end{center}
\caption{The angle of repose against cell thickness simulated
with Hookian and Hertzian Models.} \label{fig6}
\end {figure}

\begin{figure}
\begin{center}
\includegraphics[width=4 in]{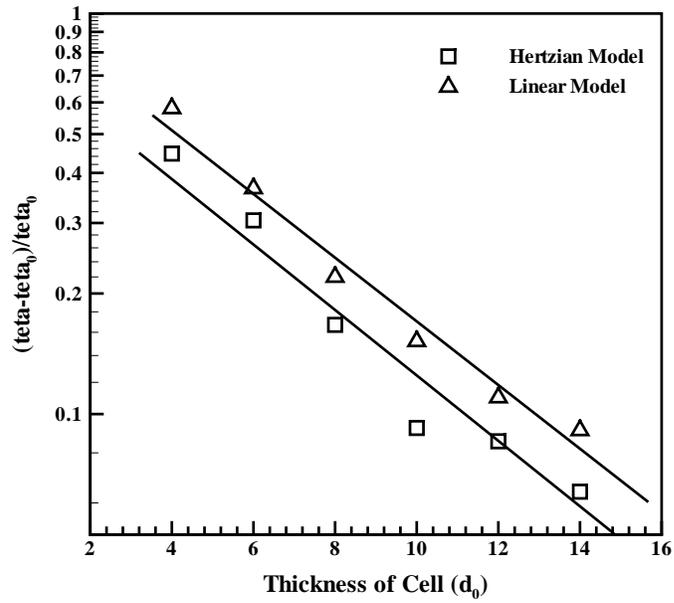}
\end{center}
\caption{Variation of $(\theta-\theta_{0})/\theta_{0}$ with cell
thickness from Hookian and Hertziam Models.} \label{fig7}
\end{figure}

To quantify the behaviour of the angle of repose when the cell
thickness is varied, simulations were performed using different
cell thickness $w$. The results presented in Fig.\ref{fig6} shows
that by increasing the cell thickness, the angle of repose,
$\theta_r(w)$, decreases and it eventually settles down to an
asymptotic value $\theta_\infty$ when the cell thickness becomes
very large compared to the size of grains. A logarithmic plot of
$(\theta_\infty-\theta_r(w))/ \theta_\infty$ versus $w$ shows
that in accordance with experimental observations, the trend
follows an exponential decay (Fig.\ref{fig7}). We also observe
that for both Hookian and Hertzian models, have the same slope,
which means that the characteristic lengths does not depend on
the model interaction. Since MD simulations on angle of repose of
granular heap formed by discharging materials have produced the
same results~\cite{xyz}, one may conclude that the observed
phenomenon is universal and does not depend in the formation
history of the pile.

\begin{figure}
\begin{center}
\includegraphics[width=4 in]{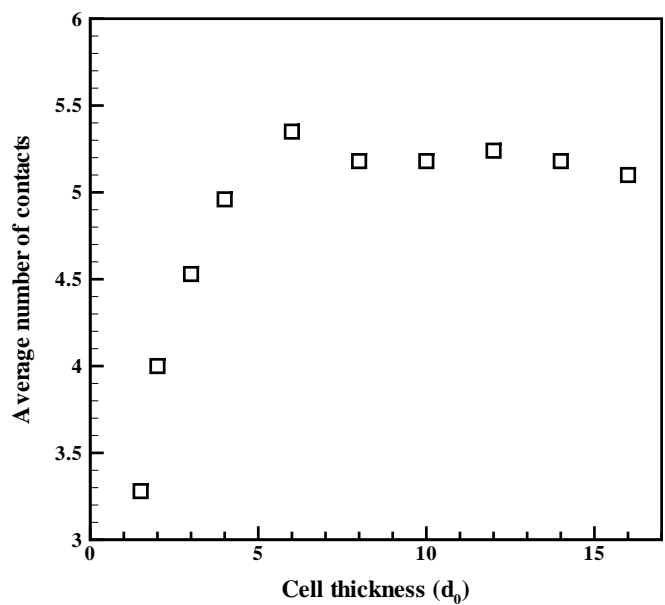}
\end{center}
\caption{The average number of contacts for interior grains
versus cell thickness from Hookian model.} \label{fig8}
\end{figure}
This phenomenon can be explained based on a dimensionality
analysis. The walls of the cell not only consume the kinetic
energy of grains via friction and inelastic collisions, but they
confine them to move in a limited space. Indeed, by changing the
cell thickness, while keeping the size of grains fixed, one can
even change the dimensionality of the space accessible to the
grains. Investigation of the average number of contacts per
grain, $\overline{z}$, in this round of simulations demonstrates
the validity of this image. In Fig.\ref{fig8} the variation of
$\overline{z}$ versus $w$ has been depicted for the interior
grains, i.e. grains which are not supported by the walls. The
fraction of interior grains decreases with $w/d$, but even at a
small value of cell thickness as $w/d=1.5$, there are still a
non-negligible fraction of such grains (about $\%8.8$ of total
grains) in the cell. We also observe that for $w>2d$, the average
number of contacts becomes less than $4$, which is the minimal
average coordination number required to obtain a static packing of
frictional spheres in three dimension~\cite{ed,sil1} indicating a
cross-over from a three dimensional regime to a two dimensional
one. At $w=1.5d$, the mean number of compact is $3.3$, very close
to the minimal value $\overline{z}=3$ needed for a stable packing
of frictional spheres in two dimension. On the other hand, for
large values of $w/d$, where the angle of repose reaches to its
bulk value, the value of $\overline{z}$ quickly reaches to the
asymptotic value $\overline{z}=5.2$. Of course, this limiting
value depends on the microscopic parameters of the granular
materials.

\subsection{Effect of particle size}

Another important and interesting parameter which we expect to
affect the piling of grains is the grain diameter $d$ itself. In
general, one expects that the angle of repose of a three
dimensional sandpile does not depend on the grain size, since  a
pile of smaller grains transforms into a pile  of larger grains
via an isotropic re-scaling of the pile
coordinations~\cite{mbsw}. Based on this idea, we expect that in
granular Hele-Shaw cells, the angle of repose of a pile to be a
function of $w/d$, because the confinement of the pile between
two vertical plates, remove the spatial symmetry in the
perpendicular direction. As such, in an isotropic re-scaling of a
sandpile in a Hele-Shaw cell, the gap between walls should be
re-scaled with grain size too.

\begin{figure}
\begin{center}
\includegraphics[width=4 in]{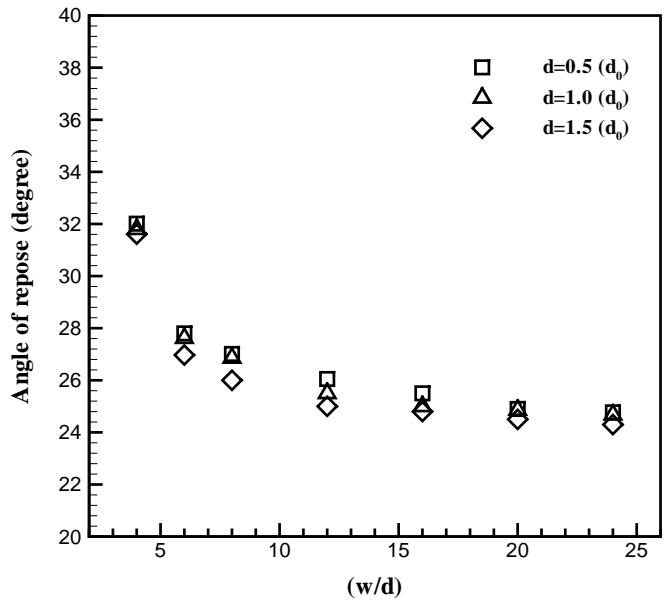}
\end{center}
\caption{Angle of repose as a function of cell thickness scaled
with particle size in Hookian and Hertzian models.} \label{fig9}
\end{figure}

In order to examine this idea, we performed a series of
simulations using the base parameter values as listed in table.I,
but with different different grain size $d$ and cell thickness
$w$, using linear model. The results are sketched in
Fig.\ref{fig9}, where the behaviour of $\theta_r$ has been
plotted against $w/d$, for several values of grain diameters. As
anticipated, all the curves almost collapse into one curve
showing that in general, the angle of repose of a sandpile in a
Hele-Shaw cell does depend on the size of grains. In the range of
parameter values we used here, the value of $\theta_r$ decays
quickly to its asymptotic value $\theta_\infty$ for $w/d>20$.
This is in agreement with our anticipation that in three
dimension the angle of repose does not depend on the size of
grains.
The data collapse in Fig.\ref{fig9} implies that the parameters
$\alpha$ and $\theta_\infty$ in Eq.(11) do not depend on the grain
size. Furthermore, the value of characteristic length should be
proportional to the grain size $d$. In the experiments conducted
by Grasselli and Herrmann, the value of $\theta_\infty$ and
$\alpha$ turn out to be size-independent too. Meanwhile, they did
not find any evidence for a linear dependence of the
characteristic length on grain diameter. In fact, the
characteristic length $l$ in their observation was the same for
all spherical grain sizes. This unexpected result could be due to
the slightly variation in the size of grains (polydispersity of
the granular materials) they used~\cite{gh}. It would also be a
result of the cohesive force between grains resulted from
capillary force for humidified grains or from van der Waals force
for dry grains less than $100 \mu m$~\cite{xyz}.

On the other hand, MD simulations on heaps formed by discharging
grains indicate that $\theta_r$ depends on particle size such
that, for a given $w/d$, larger grain have smaller angle of
repose~\cite{xyz}. This is in contrary to our results described
above. But as mentioned earlier, not only the model interactions,
but the formation histories are different in these two
simulations. The importance of the latter factor is a well-known
experimental fact. Moreover, there are evidences that in the
absence of cohesive force between grains or size-dependent sliding
coefficient~\cite{cars}, the size-dependent angle of repose could
arise as a result of introducing the rolling friction as Zhuo et
al did so~\cite{xyz}.
\subsection{Effects of frictional forces}

\begin{figure}
\begin{center}
\includegraphics[width=4 in]{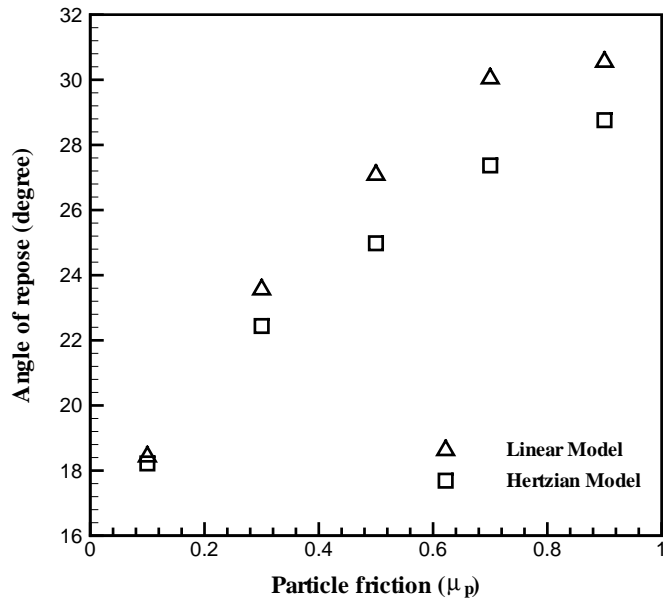}
\end{center}
\caption{Angle of repose as a function of friction coefficient
between grains from Hookian and Hertzian models.} \label{fig10}
\end{figure}

The friction between grains exhausts the translational and
rotational energy of particles. It also affects significantly the
mechanical stability of each contact in the sandpile. One expects,
therefore, that the particle-particle friction coefficient
$\mu_{p} $ be one of the key parameters in determining the angle
of repose of the pile. To investigate how this variable change
the behaviour of $\theta_r$, we performed computer simulations for
the spherical particles of fixed size with different $\mu_{p}$
using both Hertzian and Hookian models. The other values are
listed in table I. The results have been presented on
Fig.\ref{fig10} showing that for both models, the angle of repose
increases non-linearly with $\mu_{p}$. This effect is in
accordance with MD simulations of $\theta_r$ when the pile forms
from discharging materials~\cite{dis2,xyz}.

\begin{figure}
\begin{center}
\includegraphics[width=4 in]{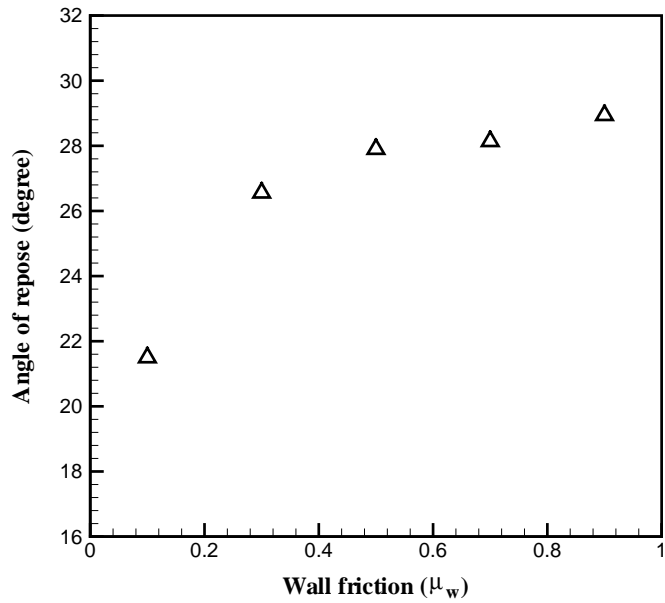}
\end{center}
\caption{Angle of repose as a function of friction coefficient
between particles and wall from Hookian model.} \label{fig11}
\end{figure}

We have also examined the effect of variation of the wall-particle
friction coefficient $\mu_w$ on $\theta_r$. This is another
important parameter which should affect the angle of repose, as
wall friction gives a torque resistance to the rotational motion
of grains. The results of simulation of the wall friction
coefficient are shown in Fig.\ref{fig11}. As indicated in the
figure, increasing the wall friction can significantly increase
the angle of repose. Similar trends have also been observed for
the grains of different sizes.

\begin{figure}
\begin{center}
\includegraphics[width=4 in]{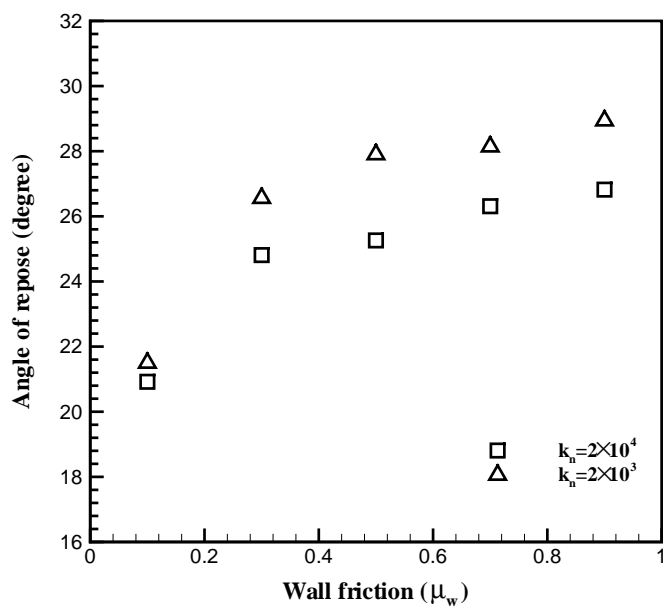}
\end{center}
\caption{Comparison of the angle of repose as a function of
friction coefficient between particles and walls from Hookian
model for two different elastic coefficients.} \label{fig12}
\end{figure}
In Fig.\ref{fig12}, we have compared the behaviour of $\theta_r$
versus $\mu _w$, for two different $k_n$ values. It is seen that
increasing material stiffness the angle of repose gets smaller,
but the change is not very significant. The decrease in
$\theta_r$ as the result of increasing $k_n$ was expected. But
the important fact is that the change in $\theta_r$ after
increasing the material stiffness coefficients by an order of
magnitude is only $2^{o}-3^{o}$. This indicates that parameters
like $k_n$ does not play a very crucial rule in formation of
sandpiles.

\section{CONCLUSIONS}

We study the dependence of the angle of repose of a sandpile
resulting from pouring granular materials in Hele-Shaw cell on
boundary conditions, particle characteristics, and the other
parameters used in the model interaction. We saw that the most
important boundary condition is the cell thickness. Increasing
this parameter induces an exponential decay in angle of repose,
in agreement with experimental findings. The same trend has been
observed in MD simulation of discharging materials, showing that
the effect is rather universal and independent of the formation
history of the pile. For cells with a large gap, the angle of
repose relaxes to an asymptotic value which should be correspond
to the angle of repose of an isotropic three dimensional heap.

At the grain level, the frictional forces are among the most
important factors. We observed that increasing particle-particle
friction and wall-particle friction coefficient increase the
angle of repose non-linearly. But the most interesting
microscopic characteristic of the single grains maybe is the
grain diameter $d$. We found that for a given cell thickness,
$\theta_r$ increases with grain size, but if the wall thickness
scales with grain diameter, the angle of repose remain unchanged.
This also implies that the angle of repose in  three dimensional
limit does not depend on the size of grains. It should be
emphasized  that these observations are just valid for the case
of cohesionless, mono-sized  grains. As such, they might be in
contrary with the result of real experiments, where these
conditions are not satisfied or the formation history is
different.

\subsection{Acknowledgments}
We would like to thank A. Amirabadizadeh, B. Haghighi and S. M.
Vaes Allaei for useful discussions. We gratefully acknowledge
IASBS parallel computer center for a generous grant of computer
time.

\end{document}